\documentclass[journal]{IEEEtran}

\usepackage{amsmath,amsfonts,bm}
\usepackage{array}
\usepackage{textcomp}
\usepackage{stfloats}
\usepackage{url}
\usepackage{verbatim}
\usepackage{graphicx}
\usepackage{cite}
\usepackage{amssymb}
\usepackage{multicol}
\usepackage[ruled,linesnumbered, noend]{algorithm2e}
\usepackage{booktabs}
\usepackage{diagbox}
\ifCLASSOPTIONcompsoc
\usepackage[caption=false, font=footnotesize, labelfont=sf, textfont=sf,subrefformat=parens]{subfig}
\else
\usepackage[caption=false, font=footnotesize,,subrefformat=parens]{subfig}

\usepackage{makecell}
\usepackage{multirow}
\usepackage{xcolor}
\usepackage{tabularx}
\usepackage{bbding}
\usepackage{threeparttable}

\begin{document}
\title{Generative AI for Advanced UAV Networking}

\author{
        Geng~Sun,
        Wenwen~Xie,        Dusit~Niyato,~\IEEEmembership{Fellow,~IEEE},
        Hongyang~Du, 
        Jiawen~Kang,
        Jing~Wu, \\
        Sumei Sun,~\IEEEmembership{Fellow,~IEEE},
        Ping Zhang,~\IEEEmembership{Fellow,~IEEE}
        \IEEEcompsocitemizethanks
        {
         \IEEEcompsocthanksitem G.~Sun is with the College of Computer Science and Technology, Jilin University, Changchun 130012, China, and also with the College of Computing and Data Science, Nanyang Technological University, Singapore 639798 (e-mail: sungeng@jlu.edu.cn).
        \IEEEcompsocthanksitem W.~Xie and and J.~Wu are with the College of Computer Science and Technology, Jilin University, Changchun 130012, China~(e-mail: xieww22@jlu.edu.cn, wujing@jlu.edu.cn)
        \IEEEcompsocthanksitem D.~Niyato and H.~Du are with the College of Computing and Data Science, Nanyang Technological University, Singapore 639798 (e-mail: dniyato@ntu.edu.sg, hongyang001@e.ntu.edu.sg).
        \IEEEcompsocthanksitem J.~Kang is with the School of Automation, Guangdong University of Technology, Guangzhou, 510006, China (e-mail: kavinkang@gdut.edu.cn)
        \IEEEcompsocthanksitem S.~Sun is with the Institute for Infocomm Research, Agency for Science, Technology and Research, Singapore 639798 (e-mail: sunsm@i2r.a-star.edu.sg)
        \IEEEcompsocthanksitem P.~Zhang is with the State Key Laboratory of Networking and Switching Technology, Beijing University of Posts and Telecommunications, Beijing 100876, China (e-mail: pzhang@bupt.edu.cn).
        }
 }
 
\maketitle

\begin{abstract}
With the impressive achievements of chatGPT and Sora, generative artificial intelligence (GAI) has received increasing attention. Not limited to the field of content generation, GAI is also widely used to solve the problems in wireless communication scenarios due to its powerful learning and generalization capabilities. Therefore, we discuss key applications of GAI in improving unmanned aerial vehicle (UAV) communication and networking performance in this article. Specifically, we first review the key technologies of GAI and the important roles of UAV networking. Then, we show how GAI can improve the communication, networking, and security performances of UAV systems. Subsequently, we propose a novel framework of GAI for advanced UAV networking, and then present a case study of UAV-enabled spectrum map estimation and transmission rate optimization based on the proposed framework to verify the effectiveness of GAI-enabled UAV systems. Finally, we discuss some important open directions.
\end{abstract}


\section{Introduction}

\par From rule-based algorithms to advanced learning models, the tasks that artificial intelligence (AI) can solve have become increasingly complex, which makes it demonstrate enormous potential for solving problems in industry, business and everyday life. Traditional AI methods, such as discriminative AI (DAI) or predictive AI (PAI), can learn special paradigms from large-scale datasets to handle classification and prediction tasks by utilizing deep neural networks. Although these AI methods provide the foundation for modern data-driven environments and demonstrate good performance in handling dynamic demands, they still face several problems, \textit{e.g.}, it relies on widely annotated datasets.

\par Fortunately, the emergence of generative AI (GAI) has alleviated the limitations faced by DAI and PAI, marking a new stage in the AI development. Specifically, GAI can learn the probability distribution from training data instead of class boundaries, and then generate trustworthy new samples based on the learned distribution. Compared to traditional AI methods, the advantages of GAI can be summarized as follows:
\begin{itemize}
    \item \textbf{Data Enhancement}: GAI has the capability to generate new data based on the learned distribution. This process can expand the training set, which helps to enhance model generalization and address dataset scarcity.
    \item \textbf{Latent Space Representation}: GAI can map the input data to the latent space during the training process, which contributes to learning the latent structure and features of the training samples. Note that this fine control is typically lacking in traditional AI methods.
    \item \textbf{Creativity}: Given the powerful generative ability of GAI and its outstanding performance in unsupervised learning, GAI has the advantages in exploratory data analysis and new field applications.
\end{itemize}

\par Benefiting from the abovementioned advantages, the importance of GAI in complex task processing has gradually emerged. In particular, the great success of ChatGPT\footnote{https://openai.com/blog/chatgpt/} and Sora\footnote{https://openai.com/sora} has ignited GAI research and spawned a multitude of applications including human-computer interaction, image processing, and video generation\footnote{https://www.coursera.org/articles/generative-ai-applications}. It is worth noting that in addition to excelling in content creation, the strong generation and exploration capabilities give it an impressive potential for handling complex communication and networking optimization problems, such as antenna array optimization~\cite{wang2023generative}. 

\par However, little research has been done on GAI for unmanned aerial vehicle (UAV) communications and networking. Currently, DAI, convex optimization, and game theory are commonly adopted for solving UAV optimization problems. However, these traditional methods may have limitations in dealing with the UAV networking problems due to the mobility of UAVs and the highly dynamic environments. Moreover, learning methods such as DAI may fail to capture the latent structure and features of the data, which results in incomplete understanding of the problem and weak ability to handle the unknown situations. 

\par GAI shows great potential for solving the issues above. Specifically, the powerful learning and generalization capabilities demonstrated by GAI can be used to optimize resource management problems in UAV networks for improving the communication performance. For example, considering the limited resources of UAVs, GAI can accurately infer the condition of the overall target area based on the data collected from part of the target area, thereby making reasonable resource allocation and trajectory planning. Although integrating GAI into UAV communications and networking offers significant advantages, there are still some issues that need to be further discussed:
\begin{itemize}
    \item \textbf{Q1}: Why is GAI suitable for UAV communications and networking?
    \item \textbf{Q2}: What UAV communication and networking issues can GAI handle?
    \item \textbf{Q3}: How does GAI handle these issues?
\end{itemize}

\par Therefore, we provide a systematic tutorial to answer the above questions. \textit{To the best of our knowledge, this is the first work to systematically demonstrate the adoption of GAI to solve UAV communication and networking optimization problems.} Our contributions are summarized as follows:
\begin{itemize}
    \item We first introduce some specific technologies and applications of GAI. Subsequently, the roles and characteristics of the UAV are demonstrated. Finally, we illustrate the limitations of DAI and briefly present GAI for UAV communications and networking.
    \item We discuss the potential of GAI to address UAV-related issues from the perspectives of communication, network, and security.
    \item We propose a novel framework for UAV communications and networking leveraging GAI. Moreover, we construct a case study to demonstrate the effectiveness of GAI for enhancing UAV-enabled spectrum sensing and communication based on the proposed framework.
\end{itemize}

\begin{figure*}[h]
    \centering
    \includegraphics[width=7.2in]{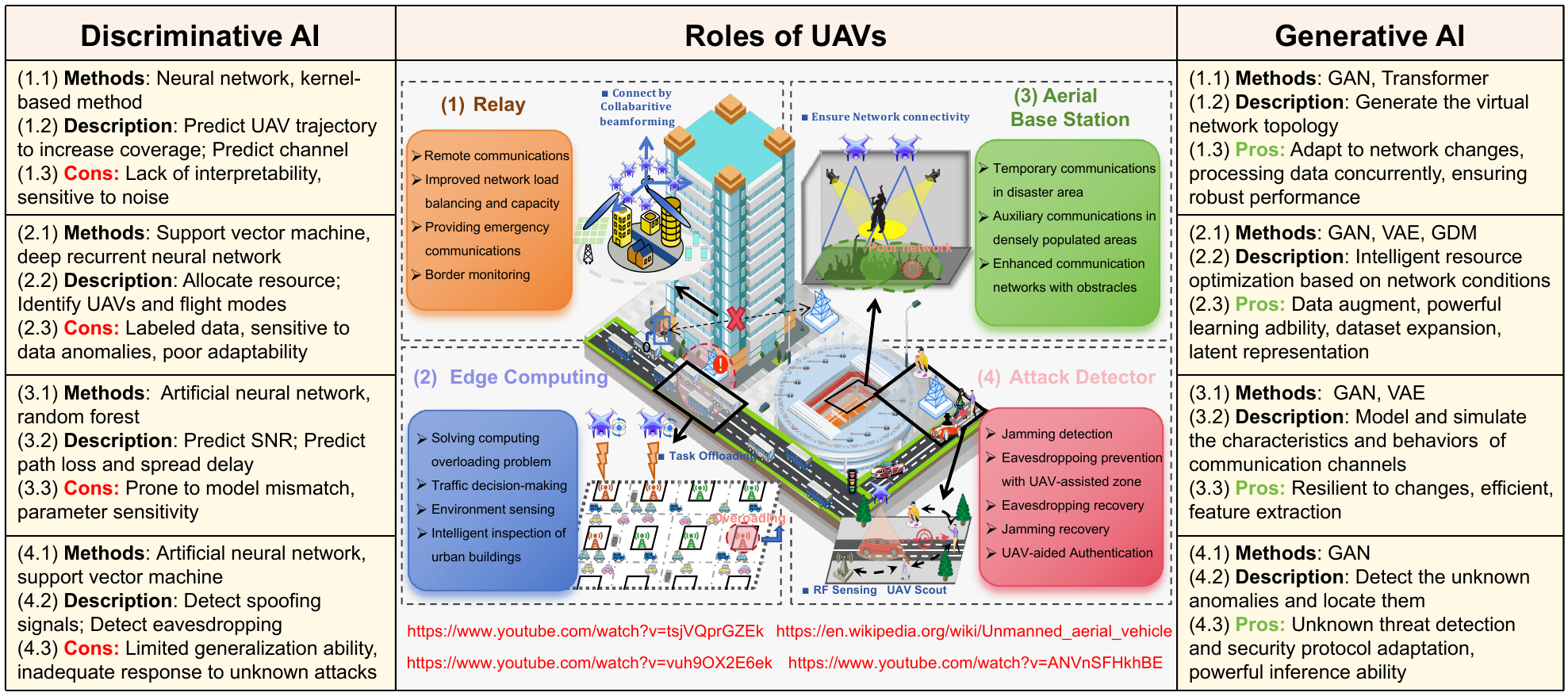}
    \caption{The roles of UAVs in communications and networking and the comparison of DAI and GAI in solving UAV optimization problem. Due to its maneuverability and computing power, the UAV can act as the aerial base station, relay and edge computing device to solve the communications and networking problems in various scenarios. Moreover, DAI and GAI are widely used to solve optimization problems in the aforementioned scenarios, where GAI stands out due to its powerful generation and learning capabilities.}
    \label{fig: Roles of UAV}
\end{figure*}

\section{Overview of GAI and UAV Networking}

\par In this section, we first introduce the key technologies and applications of GAI. Subsequently, the roles of UAV in networking are presented. Finally, we illustrate GAI on UAV in physical, network, and application layer.

\subsection{GAI and Applications}

\par GAI is based on massive general-purpose knowledge obtained from large-scale training datasets, and it can perform tasks that meet the needs of the users. Moreover, it mainly relies on the following key AI technologies:

\begin{itemize}
    \item \textbf{Large Language Model (LLM)}: LLM is trained based on a large amount of text data to learn various language patterns and structures for the purpose of understanding and generating natural language. Based on the excellent comprehension and inference generation capabilities, LLM is widely used in the fields of text generation and human-computer interaction, \textit{etc}.
    \item \textbf{Transformer}: Transformer is a sequence-to-sequence model with self-attention mechanism, which can simultaneously process information from various positions in the input sequence. Therefore, Transformer has achieved great success in natural language processing (NLP), such as machine translation and text summarization.
    \item \textbf{Generative Adversarial Network (GAN)}: GAN consists of a generator model and a discriminator model. Specifically,  The former is responsible for generating similar data to the original data, while the latter is to determine the authenticity of the data. Therefore, GANs can generate complex and realistic data by training adversarial neural networks. Currently, GAN has a wide range applications in the fields of videos and network security, \textit{etc}.
    \item \textbf{Variational Autoencoder (VAE)}: VAE is a generative model consisting of an encoder and a decoder. Specifically, the training process of VAE relies on a specific loss function that measures the difference between the reconstructed data and the original data, while considering the distribution characteristics of the latent space. Thus, VAE can learn the latent representation of the data and generate new data points similar to the training data, hence it has significant advantages in handling tasks such as signal processing and anomaly detection.
    \item \textbf{Generative Diffusion Model (GDM)}: GDM converts simple noise distributions into target data distributions through a series of reversible transformations. During the training process of the GDM, noise is gradually added to the original data, and then the inverse diffusion process is learned to construct the desired data samples from the noise. Due to the flexibility of the generation process and the high quality of the generated data, GDMs are often employed in the fields of image processing, data enhancement and recovery, and problem optimization, \textit{etc}.
\end{itemize}

\par Currently, GAI methods, which adopts the abovementioned models, have spawned a large number of applications in various fields and achieved impressive results. Next, we will briefly introduce some GAI applications from the perspective of AI-Generated Content (AIGC) and AI-Generated Everything (AIGX).

\begin{itemize}
    \item \textbf{AIGC}: AIGC refers to generating contents that meet the needs of users by GAI, mainly including the generation of media content such as text, image, video and audio. For example, LLM-based chatGPT can interact with the users and generate relevant text content according to the prompts of users, such as translations, summaries, and writing articles.
    \item \textbf{AIGX}: With the development of GAI, it has evolved to a new stage where GAI is employed to handle more complex problems in other domains and generate more complicated types of data, rather than being limited to media contents. Notably, GDM can be used in wireless network optimizations, such as maximizing the transmission rate, communication capacity, and energy efficiency.
\end{itemize}

\par The powerful problem understanding and processing capabilities that GAI have demonstrated in the abovementioned domains inspire us to apply it to solve the challenging UAV communication and networking optimization problems.

\subsection{UAV Communications and Networking}

\par UAV communication and networking systems have received increasing attention from academia, and also have already achieved significant results in practical applications~\cite{Mozaffari2019}. Specifically, several main roles of UAV systems in communication and networking 
domain are shown in Fig.~\ref{fig: Roles of UAV}, which can be detailed as follows.

\begin{itemize}
    \item \textbf{Relay}: UAVs can be utilized as mobile relay stations to connect the communication links between ground base stations and remote devices~\cite{Sun2022}. For example, in some special areas such as mountainous regions, where traditional base stations are difficult and costly to cover, UAVs can be assigned as relays to expand coverage and achieve long-distance transmission of signals. 
    \item \textbf{Aerial Base Station}: UAVs can be used as aerial base stations to provide stable and efficient communication services to ground users~\cite{Zhang}. For example, UAVs can be used as temporary communication base stations in densely populated areas to enhance communication coverage and capacity to meet the peak communication needs.

    \item \textbf{Edge Computing}: UAVs can act as edge computing devices to collect data from users for real-time analysis and processing~\cite{10382630}. For example, in intelligent traffic management, UAVs can analyze the current traffic conditions based on the obtained traffic flow statistics to make timely traffic control adjustments, which can alleviate traffic pressure.
    \item \textbf{Attack Detector}: Given their versatility, UAVs can be utilized to assist in the prevention, detection, and recovery of attacks on 5G and 6G wireless networks.
\end{itemize}

\par The widespread applications of UAV communication and networking systems are inevitable due to their advantages over traditional terrestrial systems, which are summarized as follows.

\begin{itemize}
    \item \textbf{Flexibility}: Due to their high mobility and flexible deployment capability, UAVs can move quickly to specific areas to meet the communication requirements.
    \item \textbf{Adaptability}: UAV can build temporary networks for some unexpected scenarios, supporting the temporary addition or withdrawal of UAV nodes while maintaining high availability.
    \item \textbf{High Cost-effectiveness}: Compared to terrestrial fixed systems, UAVs can carry multiple devices for providing flexible and diverse services. Moreover, the cost of UAV systems is typically lower. 
\end{itemize}

\par Note that due to the aforementioned advantages and wide range of applications, UAV communication and networking systems have been considered to play an important role in space-air-ground-sea integrated networks for 6G. Therefore, it is crucial to efficiently solve the problems faced by UAV communications and networking more efficiently.

\subsection{GAI for UAV Communications and Networking}

\par Note that DAI methods have been widely used to solve the UAV communication and networking optimization problems. For example, artificial neural network is utilized to solve the channel behavior prediction problem of UAV communications, and support vector machine is employed to solve the problem of UAV-assisted resource allocation in ultra-dense networks. However, adopting DAI for UAV networking still faces the following limitations:

\begin{itemize}
    \item \textbf{Data Dependency}: DAI methods usually rely on extensively annotated datasets and requires a large amount of well-labeled data to learn the relationships between the data. However, in the field of UAV communications and networking, obtaining large amounts of well-labeled data is difficult. Specifically, the data is often noisy in the physical layer of UAV networks, which poses a challenge for some DAI models that are sensitive to noise and vulnerable to data incompleteness.
    \item \textbf{Weak Adaptability}: DAI focuses more on patterns and features of known data, which leads to challenges when dealing with the unknown situations. In particular, the environments faced by the network layer of UAVs are often highly dynamic, hence DAI may struggle to provide flexible solutions.
    \item \textbf{Insufficient Modeling Capabilities}: DAI focuses on learning the features of data, with limited understanding of the data generation process. Thus, DAI methods are more suitable for modeling clear and simple problems. For the application layer of UAV networks, which usually involves multiple resources constraints, the data is typically large and complex. In this case, DAI may not be able to fully understand the information carried by the data for accurate modeling.
\end{itemize}

\par The limitations faced by DAI methods in solving the UAV communication and networking problems lead to the emergence of GAI. In the following, we show services that GAI can provide to UAVs from three perspectives, namely, the physical layer, network layer, and application layer. Moreover, we analyze the differences between UAV systems and other wireless systems in these services.

\subsubsection{\textbf{Physical Layer}} Compared to terrestrial communication systems, the highly dynamic and flexible nature of UAV networks makes the channel characteristics more complex. Therefore, the channel estimation requires to consider the changes of flight state in real-time for the purpose of maintaining a stable and efficient communication connection. In such cases, GAI with VAE model~\cite{xia2022generative} can improve the performance of UAV communications by generating more accurate channel parameters according to the predicted link state.

\subsubsection{\textbf{Network Layer}} In traditional terrestrial systems, the positions of nodes are typically fixed and the communication paths are relatively stable. In contrast, due to the mobility of UAVs, the network topology of UAV systems is more flexible and dynamic. In this case, GAI can generate adaptive network topology management schemes in real-time to adapt to different communication scenarios according to task requirements. For example, WaveGAN method~\cite{Odat2024} is proposed to optimize the network topology in the integration of dynamic flying ad-hoc networks.

\subsubsection{\textbf{Application Layer}} The diverse mission requirements and real-time data processing capabilities of the UAV at the application layer make it significantly different from other wireless systems. Moreover, due to the resource-constrained nature of UAVs, higher demands are placed on the process of resource allocation, and trade-off between multi-objectives, \textit{etc}. In this instance, GAI can generate intelligent resource allocation and task scheduling solutions according to the real-time demand of the current missions and environment changes, which can ensure various performances of the UAV systems. Specifically, a typical example is a GAN-based method~\cite{Li2024} to minimize the energy consumption of UAVs and task latency of ground users.

\begin{table*}[htbp]
\centering
\caption{The Use of GAI in UAV Communications and Networking.}
\label{tab: GAI for UAV}
\scriptsize
\renewcommand{\arraystretch}{1.3}
\begin{tabular}{m{1.7cm}<{\centering}|m{2.1cm}|m{2.1cm}|m{2.1cm}|m{2.1cm}|m{2.1cm}|m{3cm}}
\hline
\multirow{2}{*}{\diagbox [width=7.5em,trim=l] {\bf Issues}{\bf Models}} & \multicolumn{5}{c|}{\bf Generative AI Model} & \multicolumn{1}{c}{\multirow{2}{*}{\centering \bf Analysis}} \\
\cline{2-6}
& \makecell[c]{ \bf LLM} & \makecell[c]{ \bf Transformer} &\makecell[c]{\bf GAN} & \makecell[c]{\bf VAE} & \makecell[c]{\bf GDM} & \\
\hline
Iteractive Strategy Optimization & Learn historical relevant data~\cite{zhang2024interactive} & \centering \bf --- & \centering \bf --- & \centering \bf --- & \centering \bf --- & 
\multirow{3}{*}{\makecell[l]{Potential benefits: \\(1) More comprehensive \\ problem understanding ability \\ (2) More powerful channel \\ sensing capability \\ (3) More intelligent resource \\ allocation and more rational \\ network deployment}} \\
\cline{1-6}
Adaptive Modulation and Channel Sensing & \centering \bf --- & Predict future channels in parallel & {Train a stand-alone channel model} & Generate parameters related to channel~\cite{xia2022generative} & \centering \bf --- & \\
\cline{1-6}
Communication Resource Allocation and Optimization & \centering \bf --- & \centering \bf --- & Provide an approximation of the distributions of UAVs~\cite{Li2024} & Assign tasks to edge computing servers & Resource allocation strategy generation & \\
\hline
Route Design & \centering \bf --- & Plan UAV cluster routing & Learn 
 the routing policy~\cite{liu2022ar} & \centering \bf --- & Generate the routing strategies based on environment changes & 
\multirow{3}{*}{\makecell*[l]{Potential benefits:\\ (1) Better routing policies for \\ dynamic UAV environment\\ (2) More rational UAV network \\ topology for better performance\\ (3) Improve the accuracy of \\ network configuration}} \\
\cline{1-6}
Network Topology Design & \centering \bf --- & Plan long-term trajectory & \makecell*[l]{Generate optimized \\ network topology from \\ supervised dataset~\cite{Odat2024}} & \centering \bf --- & \centering \bf --- & \\
\cline{1-6}
Network Configuration Optimization & Analyze network traffic & \centering \bf --- & \centering \bf --- & \centering \bf --- & \centering \makecell[l]{Generate 
 the effective \\ contracts for mobile \\ AIGC services~\cite{Liu2023}} & \\
\hline
Physical Layer Security & \centering \bf --- & Predict the ideal serving beam for UAVs & Approximate mmWave channel distributions & Provide robust data transmission & Optimize beamforming and signal DoA estimation~\cite{wang2023generative} & 
\multirow{3}{*}{\makecell[l]{Potential benefits:\\ (1) Optimize UAV beamforming \\ (2) Improve the accuracy of \\ network configuration \\ (3) Enhance privacy protection \\ capability in a novel way}} \\
\cline{1-6}
Anomaly Detection & \centering \bf --- & \centering \bf --- & Detect the unknown anomalies and locate them~\cite{he2022cgan} & Detect UAV faults and anomalies & Detect network intrusion & \\
\cline{1-6}
Privacy Preservation & \centering \bf --- & \centering \bf --- & Generate training samples~\cite{zhang2021distributed} & \centering \bf --- & \centering \bf --- & \\
\hline
\end{tabular}
\end{table*}

\section{GAI-enhanced Technologies in UAV Communications and Networking}

\par In this section, we show how GAI addresses the optimization problems from UAV communication, network and security perspectives, which are summarized in Table~\ref{tab: GAI for UAV}.

\subsection{From Communication Perspective}

\subsubsection{\textbf{Interactive strategy optimization}}
\par Proper scenario simulations and correct problem formulations are cornerstones for optimizing the performance of the UAV communication and networking systems. Thus, from the design perspective, GAI can be adopted to enrich the UAV application scenarios and optimize the formulations to reduce the impact of human factors through interactive response. For example, in~\cite{zhang2024interactive}, the authors propose a problem formulation framework based on LLM and retrieval augmented generation (RAG) for a wireless communication scenario. Specifically, the users provide some detailed information that are related to the scenario (\textit{e.g.}, structure and optimization objectives) for the purpose of requiring the problem formulation. Then, the GAI agent generates the corresponding problem formulation according to the prompts of users and the knowledge database which contains some relevant extra information. The experiment results show that the problem formulation performance obtained by GAI generation outperforms manual modeling, which means that LLM with strong comprehension ability can generate accurate optimization problems through interactive processes.

\subsubsection{\textbf{Adaptive Modulation and Channel Sensing}}
\par UAVs typically perform missions in highly dynamic and constantly changing environments, which results in the methods of energy detection and cyclic detection commonly used in terrestrial-based systems being less applicable to UAV systems. In this case, we can employ GAI to analyze the collected environmental data in real-time to build channel model and adaptive modulation. For example, VAE can be used for channel sensing and modeling~\cite{xia2022generative}. Specifically, the state of the communication link is first predicted, which is subsequently input into the VAE to generate channel-related parameters such as path loss, delay, angle of arrival and angle of departure. The experiment results demonstrate this generative model outperforms standard 3GPP channel model. This demonstrates that VAE is able to capture complex statistical relationships, which is not possible in standard parametric models.

\subsubsection{\textbf{Communication Resource Allocation and Optimization}}
\par UAV communication systems require rational allocation of limited communication resources to meet the needs of different scenarios. In this context, GAI can flexibly and intelligently adjust the allocation of resources to maximize the efficiency and quality of tasks based on the real-time demands of the tasks and network conditions. For example, in~\cite{Li2024}, The authors propose a task offloading and trajectory optimization strategy based on GAN. Specifically, the energy consumption of the UAV and the task latency of the ground user are minimized by optimizing the flight trajectory, task assignment and offloading ratio. Numerical results show that this method outperforms other benchmark algorithms.

\subsection{From Network Perspective}

\subsubsection{\textbf{Route Design}}
\par The routing design of UAV networks is required to consider the factors of UAV mobility and environment dynamism, which suggests that more flexible routing algorithms and mechanisms should be adopted. We can use GAI to monitor network status and communication quality to adjust routing and forwarding policies in real-time. For example, a GAN-based method~\cite{liu2022ar} is proposed to decide the route strategy of UAV networks. Specifically, GAN is used to learn expert routing policies and combined with inverse reinforcement learning to select routes with minimum end-to-end delay based on persistent network conditions. Compared to the state-of-the-art routing protocols, this method presents outstanding performance in terms of end-to-end delay and package delivery ratio.

\subsubsection{\textbf{Network Topology Design}}
\par Due to the mobility of UAVs, the network topology is a lot more dynamic than the terrestrial communication systems. Additionally, UAV systems typically have the capability to form ad-hoc networks to support temporary communications. Therefore, intelligent network topology design is crucial for UAV systems, which contributes to improving the network performance, especially in some emergency scenarios. In this case, we can use GAI to optimize the UAV network layout and connectivity. For example, in~\cite{Odat2024}, the authors propose a GAN-based topology optimization method to maximize the network throughput in the UAV dynamic flying ad-hoc networks. In this method, GAN is adopted to learn how to generate optimized network topologies from a supervised dataset. Simulation results show that this method can find the network topology quickly, and the observed gap with the optimal topology is small in networks of different sizes.

\subsubsection{\textbf{Network Configuration Optimization}}
\par The task requirements of UAV systems have become more varied and complex, which means that UAV networks demand more intelligent and customized designs. However, traditional rule-based solutions rely on human effort and expertise, which makes them more susceptible to the human factor. In this case, GAI can be adapted to create special network designs and mechanisms according to different objectives and scenarios. For example, in~\cite{huang2023ai}, the authors propose a novel network structure called AI-Generated Network (AIGN) to generate customized network solutions to adapt to dynamically time varying environments. Moreover, the authors analyze its appealing properties, including the interpretability for intent-driven configuration.

\begin{figure*}
    \centering\includegraphics[width=\linewidth]{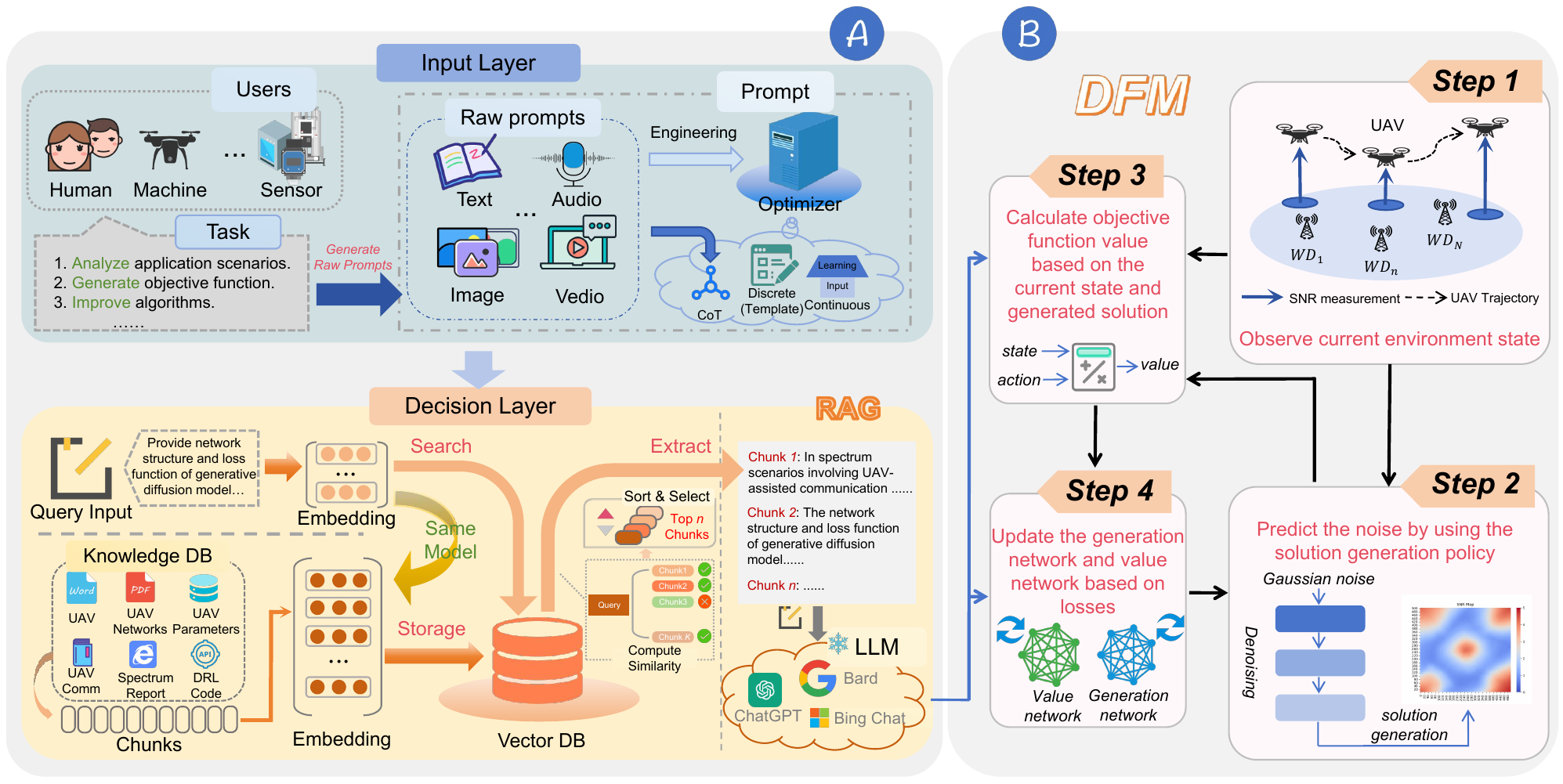}
    \caption{The framework of the proposed SEMG. In part A, the prompt optimizer generates professional prompts based on the task of users (in our case study for constructing the objective function, the network structure and loss function of the diffusion model), and subsequently the RAG is employed to output the results. In part B, the diffusion model is used to generate the solution to the optimization problem. Specifically, in Step 1, the current state is obtained. In Step 2, the diffusion model generates the solution based on the state, \textit{i.e.}, the SNR estimation. In Step 3, the objective function value is computed based on the observed state and the generated solution. In Step 4, the loss of the diffusion model network is calculated based on the objective function value and the network is updated.}
    \label{fig: Structure}
\end{figure*}

\begin{figure*}[h]
    \centering
    \includegraphics[width=7.2in]{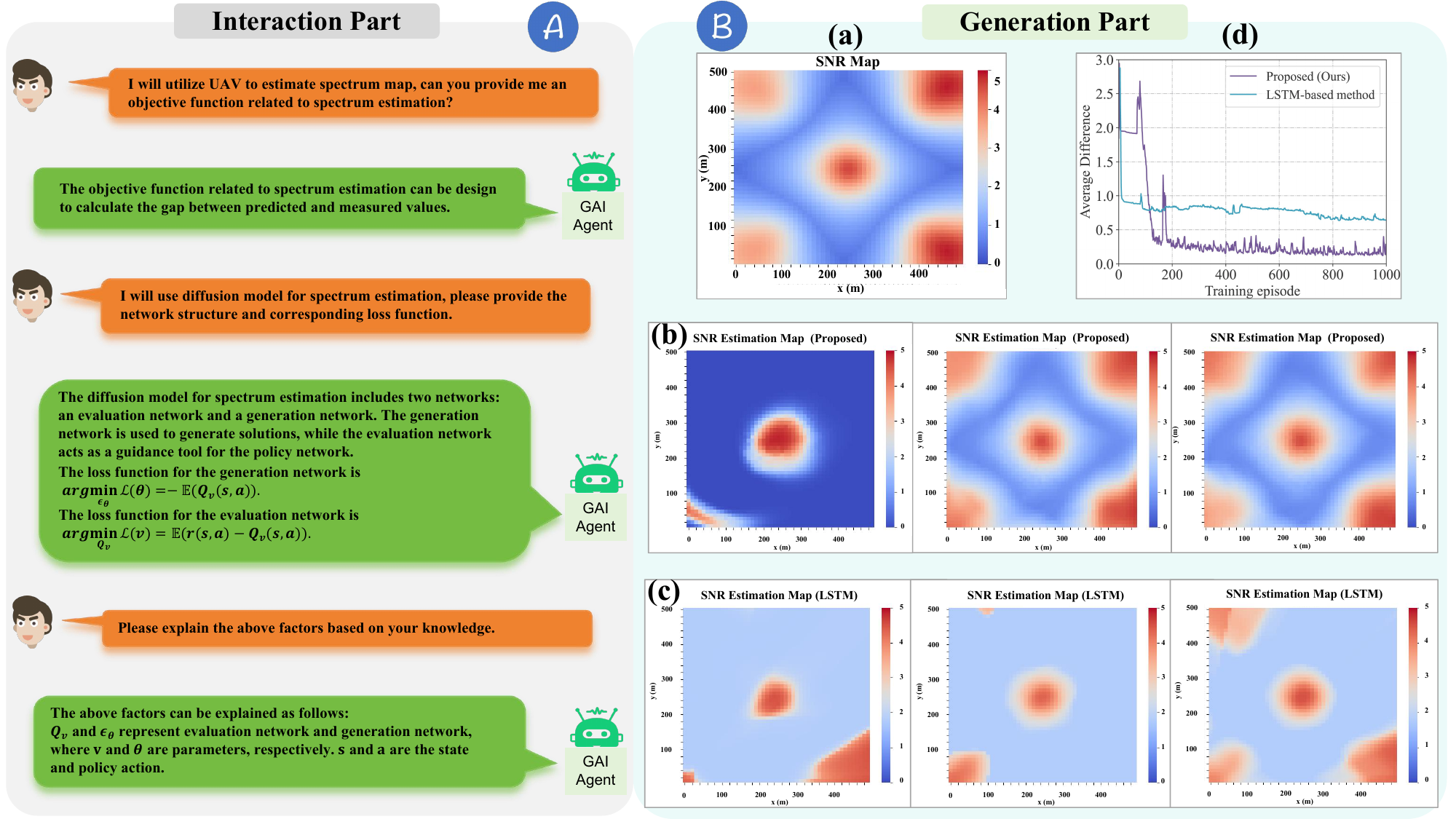}
    \caption{The experiment result of spectrum estimation. Part A shows the interaction between the user and GAI agent for the purpose of constructing the objective function, the network structure and loss function of diffusion model. Part B demonstrates the generation performance of diffusion model. (a) the true SNR map. (b) the process of generating SNR estimation map by our proposed SEMG. (c) the process of generating SNR estimation map by LSTM. (d) the difference between the measurements and estimations.}
    \label{fig: Spectrum Map Estimation}
\end{figure*}

\subsection{From Security Perspective}
\par Compared to traditional fixed terrestrial systems, the sensors and cameras carried by UAV systems may collect large amounts of sensitive data, such as flight paths and images, posing risks of privacy and information leakage during data transmission. Additionally, their high dynamicity and the broadcast nature of wireless communication increase the risks of signal variations and communication interference. Furthermore, due to resource constraints, conventional security defense measures are not applicable, which motives the tailor-made efficient security solutions for UAV systems.

\subsubsection{\textbf{Physical Layer Security}} 
\par Physical layer security which employs the natural randomness and complexity of the wireless channel is an effective and promising method to avoid eavesdroppers in UAV communications for improving the security. Beamforming or collaborative beamforming is one of the more widely used techniques in the field of UAV physical layer security. In this case, GAI can provide an optimized solution for beamforming by learning the relationship between different UAV antenna configurations and beamforming parameters. For example, the authors of~\cite{wang2023generative} propose a diffusion model-based method that efficiently estimates the direction of signal arrival under near-field conditions when the antenna spacing is more than half the wavelength. Thus, we may be inspired by this work to use GDM models to optimize beamforming vectors for UAV swarms to improve communication performance.

\subsubsection{\textbf{Anomaly Detection}}
\par UAVs are often subject to malicious nodes attacks since they are usually deployed in open environments. Therefore, network intrusion detection is essential to ensure that UAVs can safely accomplish the mission. Since GAI has strong pattern recognition and learning capabilities, it can be used to design schemes for identifying malicious nodes to secure the system. For example, in~\cite{he2022cgan}, a collaborative intrusion detection algorithm based on conditional GAN (CGAN) is proposed, in which Long Short Term Memory (LSTM) is introduced into CGAN training to improve the feature extraction ability of the network. Specifically, the data generated by CGAN is used as enhanced data for intrusion detection and classification. Compared to other models such as support vector machine, random forest and recurrent neural network, this method performs well in intrusion detection.

\subsubsection{\textbf{Privacy Preservation}}
\par UAVs may be assigned to special tasks, \textit{e.g.}, sensitive data transmission. In this case, GAI can be used to protect sensitive data transmitted or shared within UAV networks from unauthorized access or surveillance. For example, a distributed GAN framwork is proposed in~\cite{zhang2021distributed} to cooperatively model the mmWave channel for the purpose of sharing channel data in a communication-efficient and privacy-preserving approach within the UAV network. Specifically, the GAN framework trains a model to produce channel samples based on the underlying distribution of its dataset, while maintaining the confidentiality of the data distribution and without showcasing real data samples explicitly.

\section{Case Study: GAI for UAV Spectrum Map Estimation}

\par In this section, we first propose a novel framework for GAI on UAV spectrum estimation. We then introduce two case studies to demonstrate the effectiveness of the proposed framework. Specifically, the first case uses GAI to construct the spectrum estimation map based on the measurement information collected by the UAV, and the second case adopts GAI to optimize the trajectory of UAV. It is worth noting that the two case studies are interrelated and can complement each other. On one hand, the spectrum estimation map can be used by GAI to better optimize the UAV trajectory to achieve optimal transmission performance. On the other hand, the UAV can consider trajectories that can effectively collect more information to estimate spectrum accurately. The performance evaluation demonstrates this symbiotic relation between the spectrum map estimation and UAV trajectory optimization clearly.

\subsection{Motivation}
\par Spectrum cartography consists of a collection of techniques used to construct and maintain radio maps that provide useful information about radio-frequency (RF) landscape, such as received signal power and signal-to-noise ratio (SNR). As such, radio maps play an important role in solving complex tasks, such as resource allocation and mission planning. To collect the necessary measurements for building the radio maps, traditionally, technicians in a vehicle with measurement equipment would drive around the site over the past decades. With advances in mobile robotics, it is now possible to use UAVs with on-board sensors to collect the required measurement data, which is clearly more efficient in terms of time and labor costs~\cite{Zhang2021}.

\subsection{Proposed Design}
\par Utilizing data collected by UAVs for spectrum estimation and creating spectrum maps usually requires a deep understanding of wireless networks and complex professional methods. Fortunately, GAI offers a promising solution to this daunting task. Specifically, given its powerful inference capabilities, the diffusion model can be used to infer the spectrum of target area based on the data collected by the UAV, which lays the foundation for subsequent execution of more complex tasks. Note that components such as the network structure and loss design of diffusion models have a crucial impact on their optimization performance. However, manual formulation is challenging, especially for newcomers. LLM and RAG techniques can be adopted to address this issue, which can efficiently assist the users in completing the task, and significantly reduce the impact of human error. Therefore, as shown in Fig.~\ref{fig: Structure}, we propose a GAI-based UAV spectrum estimation map generator (SEMG), which contains the following components:
\begin{itemize}
    \item \textit{Interaction Part}: In the input layer, users (\textit{e.g.}, UAV operators) provide GAI with raw prompts based on task requirements, including text, images, audio and video, \textit{etc}. Then, prompt engineering is adopted to optimize the prompts, which enables a better mapping of the downstream tasks to the tasks solved by the pre-training process of the GAI agent. In the decision layer, considering that GAI cannot access real-time or non-public datasets, users can prepare some fresh or proprietary data which is relevant to their needs. Note that the prepared data above will be vectorized, indexed and stored in the vector database. Subsequently, the framework retrieves the vector database based on the query of the user and selects the $N$ data entries with the highest relevance. After that, the selected data that is regarded as the background context is integrated with the query to form the prompts, which can better guide the LLM to generate appropriate strategies that meet the requirement of users.
    \item \textit{Generation Part}: First, the current state of the environment is observed. Subsequently, the diffusion model predicts noise and generates solution based on the observed state. Finally, the objective value is calculated and the networks of diffusion model are updated according to the objective function and loss function designed by LLM in the interaction part, respectively. After training, the diffusion model can generate the corresponding spectrum estimation map based on the collected data by UAV via trained denoising network.
\end{itemize}

\par Next, We design two case studies to demonstrate the effectiveness of the proposed SEMG framework.

\subsection{Case 1: UAV-enabled Spectrum Estimation}
\subsubsection{Scenario Description}
\par The UAV is dispatched to measure SNR data in a portion of the target area. Then, the diffusion model is adopted to optimize the measurement trajectory of the UAV to more accurately predict the spectrum map of the entire target area based on limited measurement information.

\subsubsection{Performance Analysis}
\par The interaction between the user and GAI agent is presented in Part A of Fig.~\ref{fig: Spectrum Map Estimation}. As can be seen, the user intends to design the network structure and loss function of the diffusion model. Traditionally, network designers need to consult a large number of papers in order to design a reasonable model. In our proposed framework, this can be achieved automatically through the interaction between the user and the GAI agent. Note that the resulting network design perfectly aligns with the ground truth.

\par Based on the network structure and loss function designed by GAI, we can adopt diffusion model to generate spectrum estimation. Fig.~\ref{fig: Spectrum Map Estimation}(a) presents the true SNR map. Moreover, Fig.~\ref{fig: Spectrum Map Estimation}(b) and Fig.~\ref{fig: Spectrum Map Estimation}(c) shows the process of generating SNR estimation map by our proposed SEMG and LSTM, respectively. Compared to the spectrum map generated by LSTM, the diffusion model can gradually understand the meaning of the collected data during the training process and more accurately infer the SNR of the whole target area. In addition, Fig.~\ref{fig: Spectrum Map Estimation}(d) shows the estimation difference of our proposed SEMG and LSTM, where the decreasing curve shows that the gap between the estimation and measurement values decreases as the number of training episodes increases. It is obvious that our proposed SEMG outperforms LSTM. Therefore, the abovementioned results demonstrate the effectiveness of SEMG in UAV-enabled spectrum estimation.

\begin{figure}[h]
    \centering
    \includegraphics[width=3.5in]{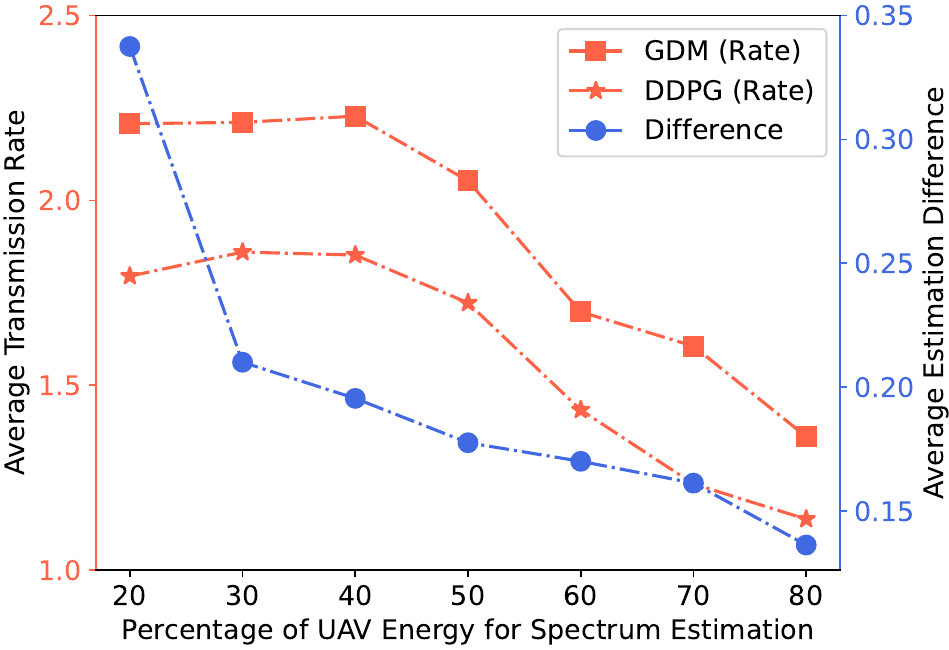}
    \caption{The impact of percentage of estimation energy on spectrum estimation difference and transmission rate.}
    \label{fig: Spectrum Map Estimation and Rate}
\end{figure}

\subsection{Case 2: UAV-enabled Joint Spectrum Estimation and Rate Optimization}
\subsubsection{Scenario Description}
\par In this scenario, the UAV acts as both spectrum estimator and data transmitter, and we aim to use diffusion model to generate the UAV exploitation trajectory for jointly achieving accurate spectrum map and high transmission rate. Considering the energy constrained nature of UAV, the impact of energy consumption on spectrum estimation and transmission performance is necessary to explore.

\subsubsection{Performance Analysis}
\par Fig.~\ref{fig: Spectrum Map Estimation and Rate} shows the impact of percentage of estimation energy consumption on spectrum estimation accuracy and transmission performance. As the energy allocated to spectrum estimation by UAV increases, the difference between the spectrum estimation map and true map gradually decreases. This is because more energy of UAV is allocated to spectrum detection will lead to the UAV collecting more information for making the spectrum map accurately. However, UAVs are energy-constrained aerial platforms, and allocating more energy for spectrum estimation means less energy are available for data transmission. Therefore, the curve of transmission rate shows a decrease tend when the energy allocated to spectrum estimation increases. It is worth noting that the quality of spectrum estimation also affects data transmission. This is because when the energy allocated to spectrum estimation is too low, the poor quality of spectrum estimation map affects the optimization data transmission rate. Note that our proposed GDM-based method outperforms DDPG, which means that it is especially suitable for the resource-limited UAV systems.

\section{Future Directions}
\par In this section, we will introduce three future directions for GAI on UAV communications and networks.
\subsection{Energy-Efficient GAI on UAVs} 
\par The inference of GAI involves complex computational processes, making it resource-intensive, particularly for energy-constrained platforms such as UAVs. Therefore, incorporating the operating costs of GAI into UAV systems is a crucial direction to ensure efficient utilization of resources for achieving optimal performance. 

\subsection{Secure GAI on UAV} 
\par The UAV wireless communications pose various attack threats due to the open channel. Although beamforming can resist eavesdropping attacks, it is powerless against data tampering by attackers. Therefore, blockchain-based data platforms are worth investigating in order to protect the data of GAI.

\subsection{Multimodal Processing on UAV} 
\par The recent success of Sora has sparked a surge in video generation technologies. It is worth noting that UAVs are extensively utilized in high-altitude video capture applications. Consequently, leveraging GAI for real-time processing of UAV videos to produce special effects or enhance quality emerges as a compelling research avenue.

\section{Conclusion}
\par In this article, we systematically introduce how GAI optimizes the UAV communications and networking. Specifically, we first introduce the fundamentals of GAI and the multiple roles of UAVs. Then, we discuss GAI on UAV from three perspectives of communication, network, and security. Subsequently, we propose a novel framework of GAI for UAV communications and networking, and then conduct a case study on UAV-enabled spectrum map estimation and transmission rate optimization to validate the effectiveness of the proposed GAI framework. Finally, three key future directions are shown that can further improve GAI on UAV systems. We hope that this article can inspire researchers to propose more GAI methods in wireless network areas such as UAV networks.

\end{document}